\documentclass[11pt]{article}
\usepackage[colorlinks=true]{hyperref}
\hypersetup{colorlinks=true,linkcolor=red,citecolor=blue,urlcolor=blue}
\usepackage{graphicx}
\usepackage{amsmath}
\usepackage{amssymb}
\usepackage{latexsym}
\usepackage{color}
\usepackage{subfigure}
\usepackage{pst-all}
\usepackage{pstricks,pst-node,pst-text,pst-3d}
\usepackage{ifpdf}
\usepackage{multimedia}
\usepackage{lmodern}
\usepackage{cite}
\usepackage{dsfont}
\usepackage{bbm}
\usepackage{lipsum}
\usepackage{url}
\usepackage{textcomp}
\usepackage{bbold}
\usepackage{epstopdf}
\usepackage{epsfig}
\usepackage{mathtools}
\usepackage{nccmath}
 \usepackage[normalem]{ulem}
\usepackage{ifpdf}
\ifpdf
\else
  \usepackage{pstricks}
\fi
  \hypersetup{pdfstartview=XYZ}
\newcommand{\bra}{\begin{array}}
\newcommand{\era}{\end{array}}
\newcommand{\beq}{\begin{equation}}
\newcommand{\eeq}{\end{equation}}
\newcommand{\beqar}{\begin{eqnarray}}
\newcommand{\eeqar}{\end{eqnarray}}

\def\BC{\bb C}
\def\_\BC{\bbi C}

\def\( {\left(}
\def\) {\right)}
\def\[ {\left[}
\def\] {\right]}
\def\no2 {{\textstyle{n\over 2}}}

\def\dag {{\dagger}}

\def \D {{\mathcal D}}

\begin{document}
\thispagestyle{empty}
\begin{center}

\vspace{1.8cm}
 \renewcommand{\thefootnote}{\fnsymbol{footnote}}


 {\Large {\bf Quantum speed limit of Jaynes-Cummings model with detuning for arbitrary initial states }}\\

\vspace{1.5cm} {\bf Yasin Shahri}$^{1}$, {\bf Maryam Hadipour}$^{1}$, {\bf Saeed Haddadi}$^{2,3}$, {\bf Hazhir Dolatkhah}$^{4}$ and\\ {\bf Soroush Haseli}$^{1,2}${\footnote { email: {\sf
soroush.haseli@uut.ac.ir}}}\\
\vspace{0.5cm}

$^{1}$ {\it  Faculty of Physics, Urmia University of Technology, Urmia, Iran}\\ [0.3em]
$^{2}$ {\it  School of Physics, Institute for Research in Fundamental Sciences (IPM), P.O. Box 19395-5531, Tehran, Iran}\\ [0.3em]
$^{3}${\it Saeed's Quantum Information Group, P.O.Box 19395-0560, Tehran, Iran}\\[0.3em]
$^{4}${\it RCQI, Institute of Physics, Slovak Academy of Sciences, \\D\'{u}bravsk\'{a} cesta 9, 84511 Bratislava, Slovakia}\\[0.3em]

\end{center}
\baselineskip=18pt
\medskip
\vspace{3cm}
\begin{abstract}
The quantum speed limit (QSL) of the Jaynes-Cummings model with detuning for arbitrary initial states is investigated. We mainly focus on the influences of the detuning, width of Lorentzian spectral density, and coherence of the initial state on the non-Markovian speedup evolution in an open system. It is found that even in the Markovian regime, increasing the detuning parameter leads to quantum speedup. Moreover, we reveal that the QSL has an inverse relation with the population of the initial excited state. Notably, we show that the QSL depends on the quantum coherence of the system's initial state such that the maximal coherent state can saturate its bound.
\end{abstract}
\vspace{1cm}

\newpage
 \renewcommand{\thefootnote}{*}
\section{Introduction}
In the study of time evolution of quantum systems, the first question that comes to mind is how fast a quantum process evolves to an orthogonal state?  Trying to find the answer of this question is the starting point to understand the concept of quantum speed limit (QSL). Actually, the QSL is the maximum  evolution speed of a quantum system. Knowing the QSL of the  evolution of a quantum system is useful and has  particular importance in various fields of quantum physics, such as quantum communication\cite{Bekenstein1981}, quantum computation\cite{Lloyd2000}, and quantum metrology\cite{Giovannetti2011}. In fact, the existence of decoherence makes the attempt to estimate the evolution time of a quantum process to be of key importance in designing quantum control protocols. In order to design quantum control protocols in the implementation of quantum information tasks, estimating the evolution time of a quantum process is of key importance.

How to speedup the evolution of quantum systems is one of the fundamental problems and questions in quantum theory \cite{Deffner2017,Hilgevoord2002,Deffner20172,Pfeifer1993,Bukov2019,Funo2019}. The shortest possible time  for a quantum system to transition from an initial state to a final state separated by a predetermined distance is known as the QSL time \cite{Hegerfeldt2013,Garcia2019,Kobe1994,Jones2010,Xu2016,Deffner2011,Shao2020,Russell2014,Hu2020}. Based on theoretical and experimental study models, QSL time has a wide range of operational applications in quantum computation and transportation \cite{Cheneau2012,del2021,Taddei2013,Escher2011,Lam2021}. Besides, the study of QSL time is of key importance in the development of quantum information technology and quantum optimal theory \cite{Poggi2019,Caneva2009,Zhang2021,Marvian2015}.  The concept of QSL time  has been  also used in the study of quantum batteries \cite{Bai2020,Campaioli2017,Hovhannisyan2013,Binder2015}.

The first definition for QSL time is provided for the  case of closed quantum systems whose their time evolution is described by the unitary time evolution operator.  In Ref. \cite{Mandelstam1991}, Mandelstam and Tamm (MT) show that the minimal time required for a closed quantum system to transform from an initial state to an orthogonal one is described by the following bound
\begin{equation}\label{MT}
\tau \geq \frac{\pi \hbar}{2 \Delta E},
\end{equation}
where $\Delta E $ is the energy variance of the system with $\Delta E ^{2} = \langle \psi \vert H^{2} \vert \psi \rangle - \langle \psi \vert H \vert \psi \rangle^2$ and $H$ is the Hamiltonian of system. The bound in Eq. (\ref{MT}) is known as MT bound. In addition, Margolus and Levitin (ML) have obtained another bound based on the average energy  as  \cite{Margolus1998}
\begin{equation}\label{E}
\tau \geq \frac{\pi \hbar}{2 E},
\end{equation}
where $E=\langle \psi \vert H \vert \psi \rangle $ is mean energy. It is also assumed  that the energy of the ground state $E_0$ is zero. The bound in Eq. (\ref{E}) is known as ML bound. By combining the MT and ML bound, a unified tight bound can be obtained for closed quantum systems as \cite{Levitin2009}
\begin{equation}
\tau \geq \max \{\frac{\pi \hbar}{2 \Delta E}, \frac{\pi \hbar}{2 E}\}.
\end{equation}
By using various metrics such as trace distance, Bures angle and relative purity, MT and ML bounds have been successfully generalized \cite{Deffner2013,Wu2018,Cai2017,del2013,Ektesabi2017,Liu2016}. In addition, for  both unitary and non-unitary evolutions, a tight bound for QSL time can be obtained by the distance between the Bloch vectors describing the state of the system \cite{Campaioli2018,Campaioli2019}. The ratio between QSL time and the actual evolution time determines whether is it possible to speedup the
dynamical evolution. If this ratio is equal to unity, then the QSL bound is saturated and there is no possibility of speeding up the
dynamical evolution. If the ratio of the QSL time to actual evolution time is less than one, then the QSL bound is not saturated and there exist the potential for dynamical evolution speedup.

A real quantum system interacts with its surroundings and the environment disturbs the system as a huge resource of decoherence and dispersion. The study of open quantum systems from various aspects has been the subject of many new researches in the field of quantum information in recent years due to their wide application in quantum information protocols \cite{Breuer2007,Alicki2007,Rivas2012}. An attractive way of dealing with open quantum systems is through the flow of information between the system and its surroundings. From the insight of memory effects, the evolution  of open quantum systems are divided into two categories: Markovian and non-Markovian. Indeed, Markovian process is known as memory-less evolution. In Markovian process the information flow from system to environment monotonically. On the other hand, non-Markovian process is recognized as with memory
process where the back-flow of information from the environment to the system will occur. In the non-Markovian process, the back-flow of information is associated with  memory effects since the back-flow makes the future states of the system depend on its past states.

In Ref. \cite{Deffner2013}, it is shown that the non-Markovian nature of the process speedup the quantum evolution.  Actually, the non-Markovian effects induce the unsaturated QSL bounds for open quantum dynamics. The mechanisms to speedup the quantum evolution through the  regulation of non-Markovianity have been studied both experimentally and theoretically \cite{Cimmarusti2015,Zhang2015,Cianciaruso2017,Sun2015,Xu2019,Xu2014}. Controlling and speeding up due to the non-Markovian feature of the quantum process is completely depend on the environmental parameters. Beside non-Markovian effects, other factors that cause the unsaturated QSL bounds should also be considered \cite{Cai2017,Teittinen2019}.

In this work, the Jaynes-Cumming model with detuning will be considered.  We will study the effects of non-Markovianity and detuning on QSL time. We first determine which environmental parameters with which values cause the non-Markovianity of quantum evolution, and then we study the effect of these parameters on the QSL time. In this study, it is shown that in addition to the non-Markovian effects, the speed of quantum evolution can be increased by adjusting the detuning parameter, even in Markovian evolution. Moreover, we  will consider the effects of detuning $\delta$ and the width of the Lorantzian spectral density $\lambda$ on QSL time. Notably, we reveal that in the case of short reservoir correlation time, the non-Markovian effects speedup quantum evolution. We also find that increasing the detuning parameter and deviation from the resonance mode will speedup the quantum process for both Markovian and non-Markovian processes.

The work is organized as follows. In Sec. \ref{model}, the considered model is introduced. The non-Markovianity for considered model is studied in Sec. \ref{non-markovian}. The QSL time for considered model will be investigated in Sec. \ref{QSLl}. Finally, we will summarize the results in Sec. \ref{con}.

\section{The model}\label{model}
Let us consider a two level system with excited state $\vert e \rangle$ and ground state $\vert g \rangle$. The system interacts with an environment consist of the quantized modes of high-Q cavity \cite{Li2010,Xu2018}. The model can be characterized by the following Hamiltonian
\begin{equation}
H=\frac{1}{2} \omega_0 \sigma_+ \sigma_- + \sum_k \omega_k b_k^{\dag}b_k+\sum_k  (g_k\sigma_+b_k+g_k^{*}\sigma_-b_k^{\dag}),
\end{equation}
where $\omega_0$ is transition frequency from excited to ground state, $\sigma_+$ and $\sigma_-$ are rising and lowering operators respectively, $\omega_k$ is the frequency of the $k$th field mode of cavity, $b_k$ and $b_k^{\dag}$ are annihilation and creation operators respectively and $g_k$ quantifies the coupling between the system and environment. We consider the case in which there exist a single excitation in the total system, so the initial state of the whole system can be expressed as
\begin{equation}
\vert \psi(0) \rangle = c_1(0) \vert e \rangle_S \vert 0 \rangle_E + \sum_k c_k(0) \vert g \rangle_S \vert 1_k \rangle_E,
\end{equation}
where $\vert 0 \rangle_E$ shows the vacuum state of the environment and $\vert 1_k \rangle_E$ is the state of environment with excitation in $k$th mode. Hence the state of total system at any time $t$ can be written as
\begin{equation}
\vert \psi(t) \rangle = c_1(t) \vert e \rangle_S \vert 0 \rangle_E + \sum_k c_k(t) \vert g \rangle_S \vert 1_k \rangle_E.
\end{equation}
By using the Schr\"{o}dinger equation, a series of differential equations for probability amplitudes $c_1(t)$ and $c_k(t)$ can be obtained as
\begin{equation}\label{c1}
\dot{c}_1(t)=-i \sum_k g_k \exp \left[i\left(\omega_0-\omega_k\right) t\right] c_k(t),
\end{equation}
\begin{equation}\label{ck}
\dot{c}_k(t)=-i g_k^* \exp \left[-i\left(\omega_0-\omega_k\right) t\right] c_1(t).
\end{equation}

It can be assumed that there are no photons in the initial state of the whole system, which means that  probability amplitude $c_k(0)$ is equal to zero. By solving Eq. (\ref{ck}) and substituting the solution into Eq. (\ref{c1}), one can obtain the following integro-differential equation as
\begin{equation}\label{integro}
\dot{c}_1(t)=-\int_0^{t} dt_1 f(t-t_1)c_1(t).
\end{equation}
In the above equation, $f(t-t_1)$ describes the correlation function, which is related to the spectral density of the environment by the following relation
\begin{equation}
f(t-t_1)=\int d\omega J(\omega)\exp [ i(\omega_0 -\omega)(t-t_1)] .
\end{equation}
So, it can be said that the exact form of the probability amplitude $c_1(t)$ completely depends on the choice of the environment's spectral density. Herein, the Lorentzian spectral density with detuning will be considered as
\begin{equation}
J(\omega)=\frac{1}{2 \pi}\frac{\gamma \lambda^{2}}{(\omega_0-\omega-\delta)^2 + \lambda^2},
\end{equation}
where $\delta =\omega_0-\omega_c$ is detuning and $\omega_c$ is the center frequency of the cavity. Notice that the effective coupling between the qubit and the environment decreases with increasing detuning. In the Lorentzian spectral density, the parameter $\lambda$ is the spectral width of the environment and is related to the correlation time of the environment as $\tau_E=\lambda^{-1}$. On the other hand, $\gamma$ is related to the time scale $\tau_S$, during which the state of the system changes through $\tau_S=\gamma^{-1}$ \cite{Breuer2007}. By employing the Lorentzian spectral density, the correlation function of the environment $f(t-t_1)$ can be obtained as
\begin{equation}
f(t-t_1)=\frac{1}{2}\gamma \lambda \exp[-(\lambda-i \delta)(t-t_1)].
\end{equation}
By substituting the above equation in Eq.(\ref{integro}) and using the Laplace transformation, the integro-differential equation (\ref{integro})  is easily solved and the probability amplitude $c_1(t)$ is obtained as
\begin{equation}
c_1(t)=c_1(0) k(t),
\end{equation}
where
\begin{equation}\label{kt}
k(t)=e^{-(\lambda - i \delta)t/2}\left[ \cosh\left( \frac{\Omega t }{2}\right)-\frac{\lambda - i \delta}{\Omega}\sinh\left( \frac{\Omega t }{2}\right) \right],
\end{equation}
with $\Omega=\sqrt{(\lambda-i \delta)^{2}-2 \gamma \lambda}$.
 Also, the dynamics of the model can be described by the  master equation having the following  form
\begin{equation}\label{master}
\frac{\partial}{\partial t}\rho(t)=\gamma(t) \left( \sigma_ \rho(t) \sigma_+ - \frac{1}{2}\lbrace \sigma_+ \sigma_-,\rho(t)\rbrace \right) ,
\end{equation}
where $\gamma(t)$ is the time-dependent decay rate of the model which is defined as
\begin{equation}
\gamma(t)=Re\left( \frac{2 \gamma \lambda \sinh(\Omega t /2)}{\Omega  \cosh(\Omega t /2)+(\lambda - i \delta)\sinh(\Omega t /2)}\right),
\end{equation}
in which  $Re(o)$ is the real part of $o$. The initial state of the system can be described by $\rho(0)=\sum_{i,j=1}^{2} \rho_{ij}\vert i \rangle \langle j \vert$. By solving the Eq. (\ref{master}), the state of the system at time $t$ can be obtained as
\begin{equation}\label{timedependent}
\rho(t)=\Lambda_t\rho(0)=\left(\begin{array}{cccc}
\rho_{11} \vert k(t) \vert^{2} & \rho_{12}k(t)  \\
\rho_{21}k^{*}(t) & 1-\rho_{11} \vert k(t) \vert^{2}  \\
\end{array}\right),
\end{equation}
where $\Lambda_t$ is known as quantum dynamical map, which maps initial state at time $t=0$ to the state at time $t$. The concept of quantum dynamical map will be described in the following section. Also, the population of excited states at time $t$ can be defined as $P_t=\rho_{11} \vert k(t) \vert^{2}$.

\section{Measuring non-Markovianity}\label{non-markovian}
In the theory of open quantum systems, the evolution of a quantum system can be described by a quantum dynamical map. The dynamics of an open quantum system can be described by a master equation. Let us suppose that the dynamical map is defined by a master equation in Lindblad form
\begin{equation}
\frac{\partial}{\partial t}\rho(t)=\mathcal{L}\rho(t),
\end{equation}
where $\mathcal{L}$ is the Lindbladian super-operator given as \cite{Gorini1976,Lindblad1976}
\begin{equation}\label{msr}
\mathcal{L} \hat{\blacklozenge}=-i [\hat{H},\hat{\blacklozenge}]+\sum_i \gamma_i \left[\hat{A}_i \hat{\blacklozenge} \hat{A}_i^{\dag}-\frac{1}{2} \lbrace \hat{A}_i^{\dag}\hat{A}_i, \hat{\blacklozenge}\rbrace \right],
\end{equation}
where $\hat{H}$ is the Hamiltonian of the considered system, $\gamma_i$ is decay rate and $\hat{A}_i$'s are Lindblad operators. If the Lindblad operators and decay rates  are time-independent and  decay rates have positive value, then Eq.(\ref{msr}) leads to a completely positive trace-preserving (CPTP) map $\Lambda(t,0)=\exp[\mathcal{L}t]$. For All $t_1,t_2 \geq 0 $, the CPTP map satisfies the semi-group property as
\begin{equation}
\Lambda(t_1+t_2,0)=\Lambda(t_1,0)\Lambda(t_2,0).
\end{equation}
In this situation, the dynamical CPTP map describes a conventional Markovian process. Of course, it is possible that Hamiltonian $\hat{H}$, the Lindblad operators $\hat{A}_i$ and decay rates have an explicit dependence on time. In such a situation, Eq.(\ref{msr}) describes a time-dependent Markovian process if the value of decay rate is positive at all times, i.e., $\gamma_i(t) \geq 0$. In the time-dependent scenario, the dynamical map  can be written as
\begin{equation}
\Lambda(t,t_0)= T_{\longleftarrow} \exp\left[ i \int_{t_0}^t \mathcal{L}(s)ds\right] ,
\end{equation}
where $T_{\longleftarrow}$ is time ordering operator \cite{haddadiPRE}. The Markovian quantum dynamical maps $\Lambda(t,0)$ have special property that they satisfy the divisibility condition. The divisibility condition is defined in such a way that a  CPTP map can be expressed as a composition of two other CPTP maps as
\begin{equation}
\Lambda(t_2,t_0)=\Lambda(t_2,t_1)\Lambda(t_1,t_0).
\end{equation}
\begin{figure}
\centering
    \includegraphics[width =0.58 \linewidth]{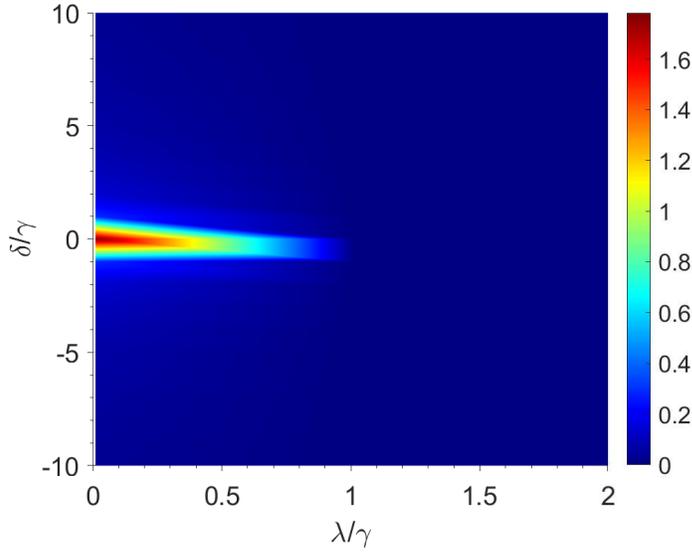}
    \caption{Non-Markovianity $\mathcal{N}$ as functions of detuning parameter $\delta$ and spectral width of the environment $\lambda$.}
    \label{Fig1}
  \end{figure}
It is important to mention that the value of decay rate $\gamma_i(t)$ may become negative in some time intervals during the evolution. In such a situation, there is an intermediate dynamical map $\Lambda(t_2,t_1)$ which is not CPTP in the interval that $\gamma_i(t)$ is negative. So, the divisibility condition violated in these intervals \cite{Laine2010,Breuer2012}. It is recalled that what has been introduced as a time-dependent Markovian process is based on the divisibility property. Many criteria for measuring the non-Markovianity of the quantum process are based on the divisibility condition.  However, some   criteria are based on the back-flow of information from the environment to the system. In this work, we focus on a criterion that is based on the flow of information between the system and the environment. The employed measure is quantified by trace distance of a pair of arbitrary states $\rho_1(t)$ and $\rho_2(t)$, which is given by \cite{Laine2010,Breuer2012}
\begin{equation}\label{TD}
D(\rho_1(t),\rho_2(t))=\frac{1}{2}\parallel \rho_1(t) - \rho_2(t) \parallel.
\end{equation}
Actually, trace distance is  a measure to quantify the distinguishability of a pair of quantum states. So, its changes during evolution can be interpreted as the exchange of information between the system and the environment. If the the distinguishability of the two states in Eq. (\ref{TD}) decreases monotonically ( $W_t=\partial_t D(\rho_1(t), \rho_2(t))<0 $), then it can be said that information continuously flows from system to environment, however, the back-flow of information to the system will not occur and the process is Markovian. On the opposite side, $W_t >0$ means that there exist back-flow of information from environment to system and so the process is non-Markovian. Accordingly, Breuer et al. \cite{Breuer2009} proposed a non-Markovian measure as
\begin{equation}\label{Bru}
\mathcal{N}(\Lambda)=\max_{\rho_1(0), \rho_2(0)}\int_{W_t >0}W_t dt.
\end{equation}
Based on Eq. (\ref{Bru}), it should be necessary to perform an optimization process on all possible initial states $\rho_1(0)$ and $\rho_2(0)$ to determine the degree of non-Markovianity.  In Ref. \cite{Breuer2009}, it is shown that the maximum value of Eq. (\ref{Bru}) will be obtained for  initial states $\rho_1(0)=\vert 0 \rangle \langle 0 \vert $ and $\rho_1(0)=\vert 1 \rangle \langle 1 \vert $, by considering a large sample set of pairs of initial states and using strong numerical evidence. By choosing these two initial  states, the trace distance in Eq. (\ref{TD}) for the model defined in this work in the previous section is obtained as
\begin{equation}\label{tr}
D(\rho_1(t),\rho_2(t))=\vert k(t) \vert^{2},
\end{equation}
where $k(t)$ is given by Eq. (\ref{kt}).

To illustrate this result, Fig. (\ref{Fig1}) shows the non-Markovianity $\mathcal{N}$ as a function of detuning parameter $\delta$ and spectral width of the environment $\lambda$. From this figure, one can see that in the resonance case when $\delta=0$, the non-Markovianity has its maximum value for small values of $\lambda$. While for the detuning case $\delta \neq 0$, the degree of non-Markovinity decreases with increasing the absolute values of detuning parameter $\delta$. In the following, we will use the results extracted from Fig. (\ref{Fig1}) to study the QSL time for both Markovian and non-Markovian cases.

\section{QSL time}\label{QSLl}
In recent works, various methods have been proposed to study the QSL time. The geometric approach in quantum information theory is usually applied to introduce a criterion for quantifying  the QSL time of quantum evolution. In Ref. \cite{Wu2018},  the authors have provided the QSL time for mixed initial states. Here, we use the method which has introduced by them. In their work, the function of relative purity has been used to define the QSL time, which is given by
\begin{equation}\label{metric}
\Theta(\rho(0),\rho(t))=\arccos\left( \sqrt{\frac{tr[\rho(0)\rho(t)]}{tr[\rho(0)^2]}}\right).
\end{equation}
Based on the metric introduced in Eq. (\ref{metric}), a comprehensive expression of QSL time for the mixed initial states is obtained as follows
\begin{equation}\label{QSL}
\tau_{QSL}=\max \left\lbrace \frac{1}{\Phi_\tau^{op}},\frac{1}{\Phi_\tau^{tr}},\frac{1}{\Phi_\tau^{hs}} \right\rbrace \sin^{2}[\Theta(\rho(0),\rho(\tau))]tr[\rho(0)^2],
\end{equation}
where $\tau$ is the actual driving time and
\begin{eqnarray}
\Phi_\tau^{op}&=&\frac{1}{\tau}\int_0^\tau dt \parallel \dot{\rho}(t) \parallel_{op},\nonumber\\
\Phi_\tau^{tr}&=&\frac{1}{\tau}\int_0^\tau dt \parallel \dot{\rho}(t) \parallel_{tr},\nonumber \\
\Phi_\tau^{hs}&=&\frac{1}{\tau}\int_0^\tau dt \parallel \dot{\rho}(t) \parallel_{hs},
\end{eqnarray}
where $\parallel \dot{\rho}(t) \parallel_{op}=\lambda_1$ is the operator norm of $\dot{\rho}(t)$ ($\lambda_1$ is the largest singular value of $\dot{\rho}(t)$), $\parallel \dot{\rho}(t) \parallel_{tr}=\sum_i \lambda_i$ is the trace norm of $\dot{\rho}(t)$($\lambda_i$'s are the all singular values of $\dot{\rho}(t)$) and $\parallel \dot{\rho}(t) \parallel_{hs}=\sqrt{\sum_i \lambda_i^{2}}$ is the Hilbert-Schmidth  norm of $\dot{\rho}(t)$. In Eq. (\ref{QSL}), if the denominator of the fraction is $\Phi_\tau^{op}$ and $\Phi_\tau^{tr}$, we have generalized ML type QSL bound for open quantum systems, while if it is $\Phi_\tau^{hs}$,  we have MT type bound on the QSL time for non-unitary dynamics.

For a matrix like $B$, the following inequality holds for the norms which are used in Eq. (\ref{QSL})
\begin{equation}\label{ineq}
\parallel B \parallel_{tr} \geq \parallel B \parallel_{hs} \geq \parallel B \parallel_{op}.
\end{equation}
From Eq. (\ref{ineq}), one can obtain the following order for denominator in Eq. (\ref{QSL}) as
\begin{equation}
\Phi_\tau^{op} \leq \Phi_\tau^{hs} \leq \Phi_\tau^{tr}.
\end{equation}
It is clear from above equation that  ML type bound based on the operator norm is the tightest QSL time bound for non-unitary evolution.

Now, we would like to use this bound to investigate the effects of environmental parameter on QSL time for the model which has described in Sec. \ref{model}. According to the fact that the bound can be used for the mixed initial states, let us consider a general two-level system. In the Bloch representation, the general initial two-level state can be written as
\begin{equation}
\rho(0)=\frac{1}{2}\left(\begin{array}{cc}
1+r_z  & r_x-i r_y  \\
r_x + i r_y & 1-r_z   \\
\end{array}\right),
\end{equation}
where $r_x$, $r_y$ and $r_z$ are components of the Bloch vector $\textbf{r}$. So,
from Eq. (\ref{timedependent}), the density matrix at time $t$ can be written as
\begin{equation}\label{density}
\rho(t)=\frac{1}{2}\left(\begin{array}{cc}
(1+r_z)\vert k(t)\vert^{2}  & (r_x-i r_y)k(t) \\
(r_x + i r_y)k^{*}(t) & 2-(1+r_z)\vert k(t)\vert^{2}   \\
\end{array}\right).
\end{equation}
As mentioned before, the ML type bound of QSL time based on the operator norm is the tightest bound, therefore, the QSL time can be considered as
\begin{equation}\label{QSLop}
\tau_{QSL}= \frac{1}{\Phi_\tau^{op}} \sin^{2}[\Theta(\rho(0),\rho(\tau))]tr[\rho(0)^2],
\end{equation}
From Eq. (\ref{density}) and using Bloch representation, the QSL time is obtained as
\begin{equation}\label{QSLop2}
\tau_{QSL}=\frac{(1-k(\tau))[r_x^{2}+r_y^{2}+r_z(1+r_z)(1+k(\tau))]}{\frac{1}{\tau}\int_0^{\tau} \vert \dot{k}(t) \sqrt{r_x^2+r_y^2+4 k(t)^{2}(1+r_z)^{2}}\vert dt}.
\end{equation}
\begin{figure}
    \centering
    \includegraphics[width =0.58 \linewidth]{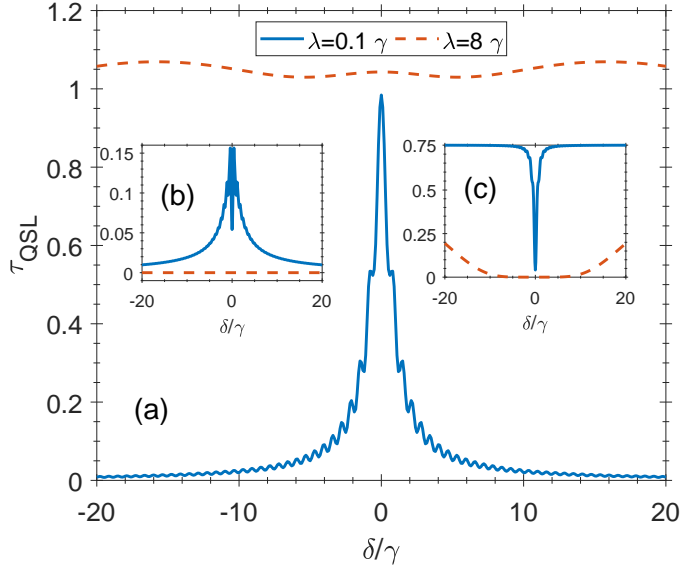}
    \centering
    \caption{ QSL time $\tau_{QSL}$ (a),  non-Markovianity $\mathcal{N}$ (b), and population of excited states $P_\tau$  (c) versus $\delta/\gamma$ for both Markovian ($\lambda=8\gamma$) and non-Markovian ($\lambda=0.1 \gamma$) regimes with $r_x=r_y=r_z=0.5$ at $\tau=10$.}
    \label{Fig2}
  \end{figure}
In Fig. \ref{Fig2}a, the QSL time is plotted versus detuning parameter $\delta$.  Two cases will be considered: Markovian and non-Markovian dynamics. From Fig. \ref{Fig1}, one can choose $\lambda=8 \gamma $ and $\lambda=0.1 \gamma$ to have the Markovian and non-Markovian dynamics, respectively. For both Markovian and non-Markovian dynamics dynamics, the actual driving time has considered to be $\tau=10$. It can be seen that the QSL time for non-Markovian dynamics is shorter than Markovian case. So, one can conclude that the non-Markovian effects can speedup the quantum evolution. Fig. \ref{Fig2}b shows the degree of non-Markovianity during actual evolution time versus detuning parameter. As can be seen for $\lambda=8 \gamma$, the the evolution is Markovian while for $\lambda=0.1 \gamma$ we have non-Markovian dynamics. It is also observed that the degree of non-Markovianity decreases with increasing detuning parameter, in agreement with Fig. \ref{Fig1}. Besides, the population of excited states $P_\tau$ is plotted versus detuning parameter in Fig. \ref{Fig2}c. As can be seen, the population of excited states increases with increasing detuning parameter for both markovian and non-Markovian dynamics. So, it can be concluded that there exists inverse relation between the population of excited states and the QSL time.
  \begin{figure}
    \centering
    \includegraphics[width =0.58 \linewidth]{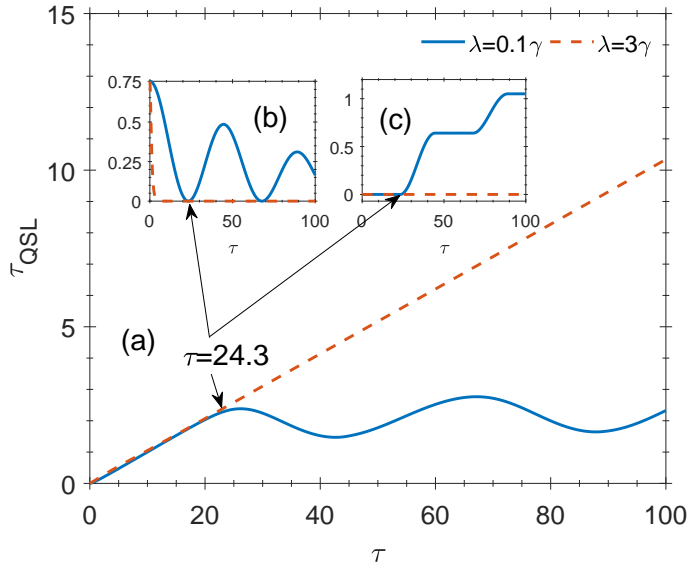}
    \centering
    \caption{QSL time (a),  population of excited states  (b), and non-Markovianity (c) versus the actual evolution time $\tau$ for both Markovian ($\lambda=3\gamma$) and non-Markovian ($\lambda=0.1 \gamma$) regimes with $r_x=r_y=r_z=0.5$ at resonance case $\delta=0$.}
    \label{Fig3}
  \end{figure}

Now, in Fig. \ref{Fig3}a, we illustrate the QSL time as a function of actual driving time $\tau$  in both Markovian ($\lambda=3 \gamma$) and non-Markovian ($\lambda=0.1$) regimes  for $\delta=0$. As expected, the QSL time for non-Markovian regime is shorter than that of the Markovian regime. Fig. \ref{Fig3}b shows the time variation of the population of excited states $P_\tau$ for both Markovian and non-Markovian regimes. By comparing Fig. \ref{Fig3}a and Fig. \ref{Fig3}b, it can be seen that  there is an inverse relation between the QSL time and the population of excited states $P_\tau$. In Fig. \ref{Fig3}c, the degree of non-Markovianity  is plotted in terms of actual evolution time $\tau$. As expected for $\lambda=3 \gamma$, non-Markovianity is zero for all actual time while for $\lambda=0.1$ at $\tau=24.3$, the non-Markovian nature of the evolution is revealed. The interesting point in Fig. \ref{Fig3} is that with the revelation of the non-Markovian nature of the evolution at $\tau=24.3$, the fluctuations of population of excited states will be started and the QSL time will be shorter than the QSL time in Markovian regime.
  \begin{figure}
    \centering
    \includegraphics[width =0.58 \linewidth]{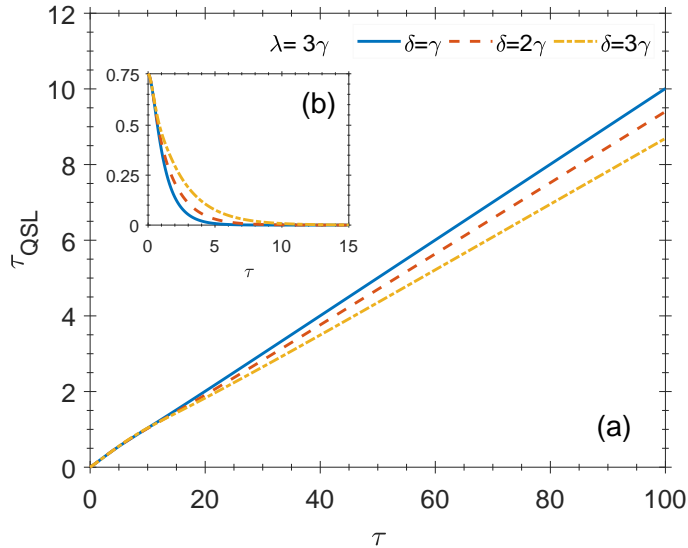}
    \centering
    \caption{ QSL time (a) and population of excited states (b) in terms of actual evolution time $\tau$ for different values of detuning parameter $\delta$ in Markovian regime with $\lambda=3 \gamma$ and $r_x=r_y=r_z=0.5$. }
    \label{Fig4}
  \end{figure}

In Fig. \ref{Fig4}a, the QSL time is plotted versus actual driving time $\tau$ for different values of detuning parameter $\delta$ in Markovian regime with $\lambda=3 \gamma$. From this plot, it can be seen that the QSL time decreases with increasing detuning parameter $\delta$ in Markovian regime. It is an interesting result because in the absence of non-Markovian effects and just by adjusting the detuning parameter $\delta$, the evolution speed has been increased. On the other hand, Fig. \ref{Fig4}b represents the population of excited states $P_\tau$ versus $\tau$  for the same values in Fig. \ref{Fig4}a. From Fig. \ref{Fig4}b, we see that the population of excited states increases with increasing the detuning parameter.

\begin{figure}
    \centering
    \includegraphics[width =0.58 \linewidth]{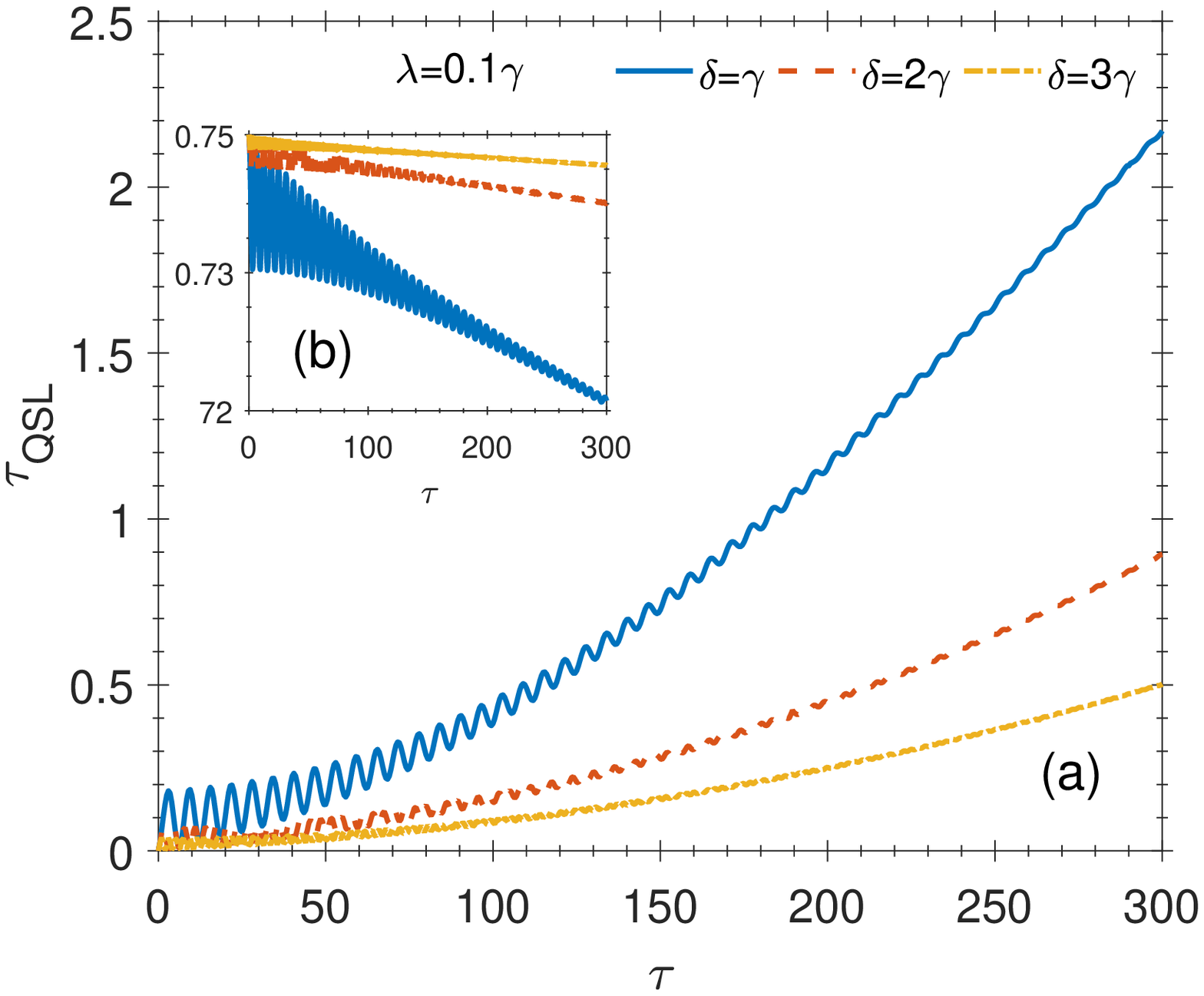}
    \centering
    \caption{ QSL time (a) and population of excited states (b) in terms of actual evolution time $\tau$ for different value of detuning parameter $\delta$ in non-Markovian regime with $\lambda=0.1 \gamma$ and $r_x=r_y=r_z=0.5$.}
    \label{Fig5}
  \end{figure}

Fig. \ref{Fig5} is similar to Fig. \ref{Fig4}, with the difference that the QSL time and the population of excited states  have been shown in a non-Markovian regime. In Fig. \ref{Fig5}a, it can be observed that for the non-Markovian regime, we have quantum speedup  with increasing the detuning parameter.  Here, although the non-Markovian effects themselves are effective for protecting the population of excited states, one can notice that the increase in the detuning parameter $\delta$ still plays an important role in protecting the population.

Notably, by using the $l_1$-norm-based measure to quantify the quantum coherence \cite{Baumgratz2014,Yu2009}, it can be concluded that the coherence of the initial state $\rho_0$ is obtained as $C(\rho_0)=\sqrt{r_x^2+r_y^2}$. Also, the $z$ component of the Bloch vector $r_z$ can be considered as the population of initial excited state, which we denote it by $\langle \sigma_z \rangle_0$. Considering these cases, Eq. (\ref{QSLop2}) can be rewritten in terms of the coherence of the initial state and the population of initial excited state as
\begin{equation}
\tau_{QSL}=\frac{(1-k(\tau))[C(\rho_0)^{2}+\langle \sigma_z \rangle_0(1+\langle \sigma_z \rangle_0)(1+k(\tau))]}{\frac{1}{\tau}\int_0^{\tau}\vert \dot{k}(t) \sqrt{C(\rho_0)^{2}+4k(t)^{2}(1+\langle \sigma_z \rangle_0)^{2}}\vert dt}.
\end{equation}
   \begin{figure}
    \centering
    \includegraphics[width =0.47 \linewidth]{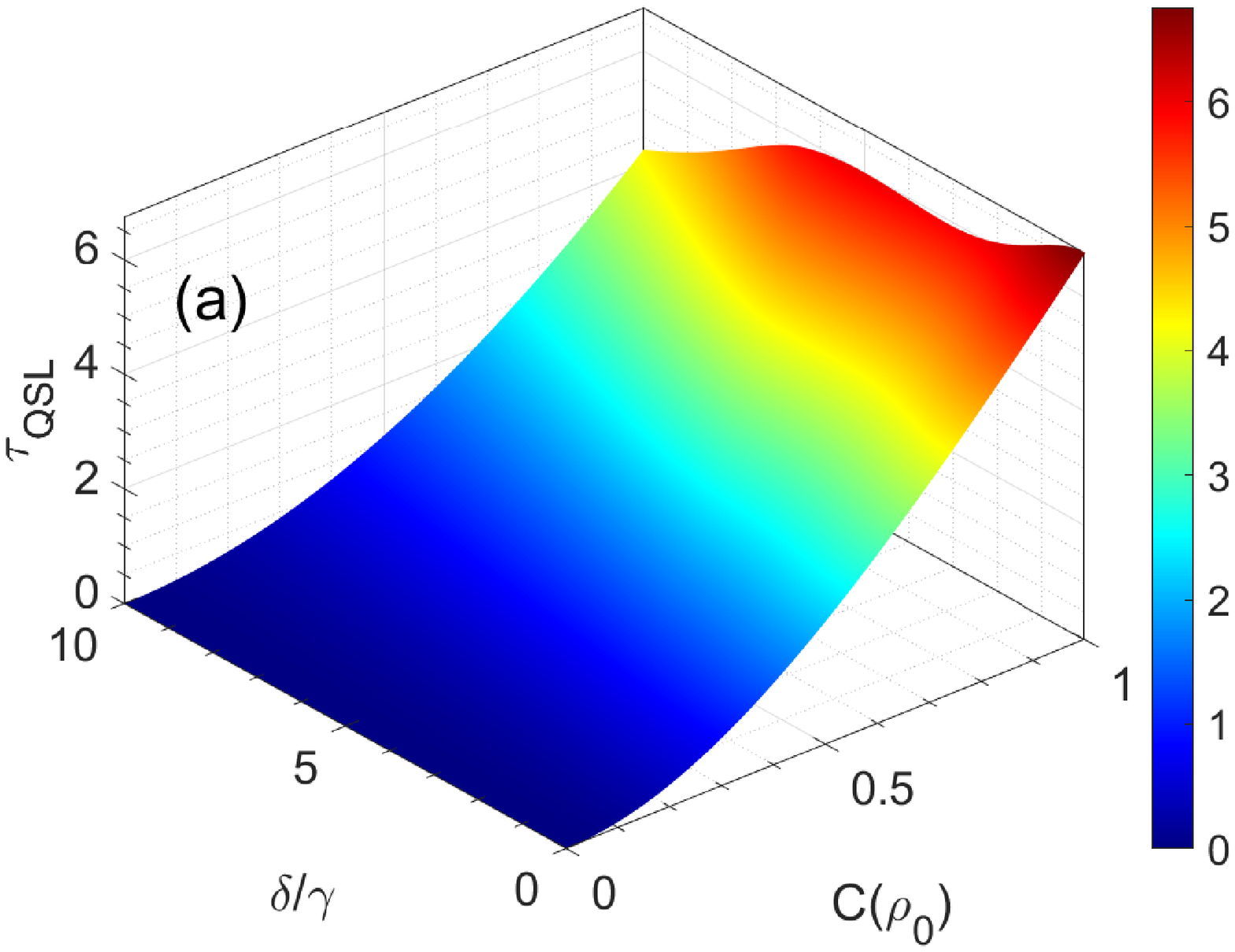}\quad
    \includegraphics[width = 0.47\linewidth]{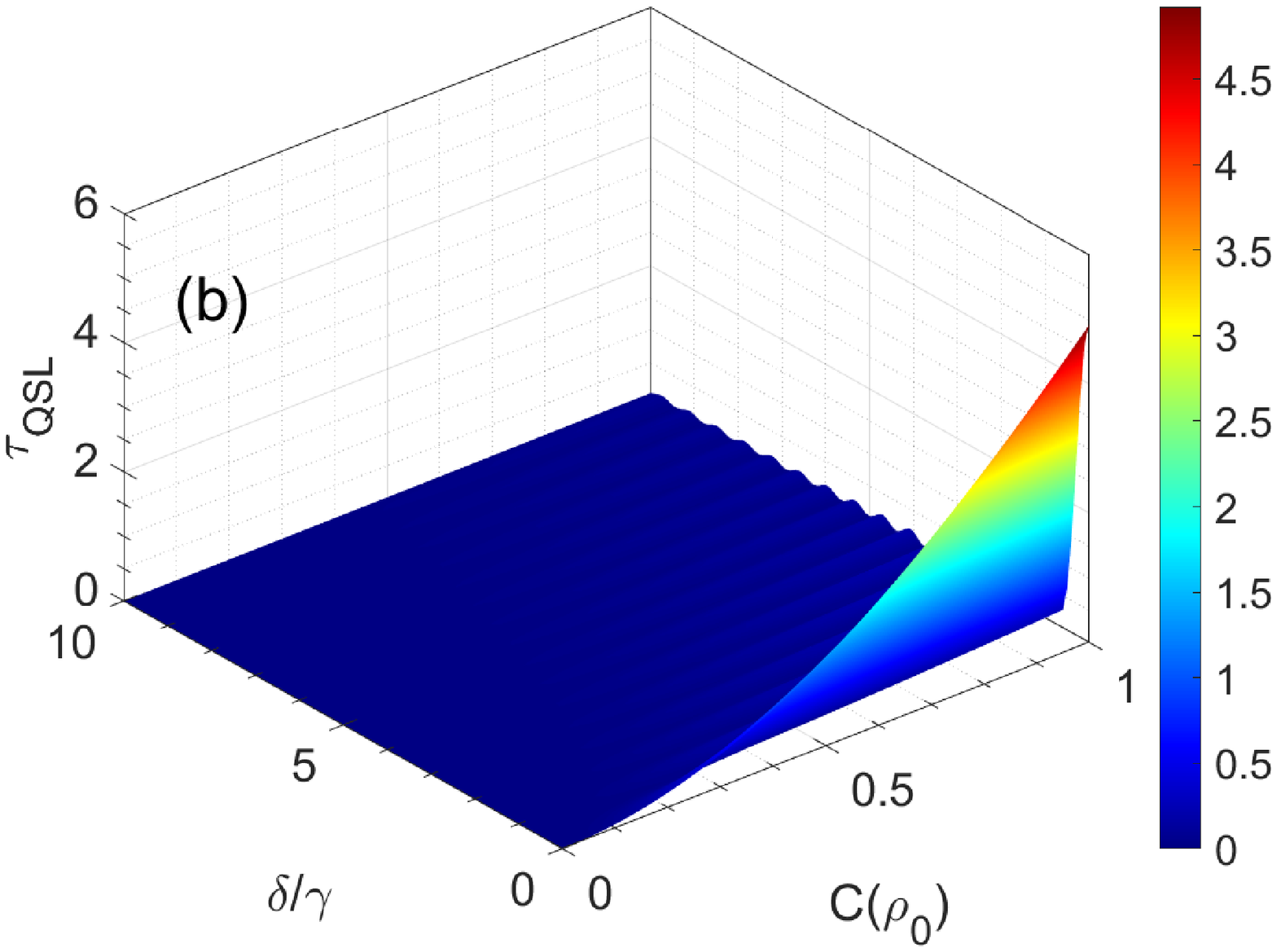}
    \centering
    \caption{QSL time as  functions of quantum coherence of initial state $C(\rho_0)$ and the detuning parameter $\delta$ when population of initial excited state is $\langle \sigma_z \rangle_0=0$ at $\tau=10$. (a) Markovian dynamics with $\lambda=3\gamma$  and (b) non-Markovian dynamics with $\lambda=0.1\gamma$.
}
\label{Fig6}
  \end{figure}

In Fig. \ref{Fig6}, the QSL time has been shown as  function of the quantum coherence of initial state $C(\rho_0)$ and detuning parameter for both Markovian and non-Markovian regimes. From Fig. \ref{Fig6}a, we see that in the Markovian regime, the QSL time has direct relation with the coherence of initial state such a way that with increasing the coherence of initial state, the QSL time increases. Besides, it is observed that the QSL time decreases with increasing the value of the detuning parameter. In a similar way, we have same interpretation for Fig. \ref{Fig6}b with the difference that the dynamics is non-Markovian and the QSL time is shorter than Markovian regime in Fig. \ref{Fig6}a.

\section{Conclusion}\label{con}
In this work, the quantum speedup mechanisms involving the regulation of detuning in Jaynes-Cumming model are investigated. Here, the relative purity based quantum speed limit bound is selected to study the quantum speedup in Jaynes-Cumming model with detuning.  We have first determined the range of environmental parameters in which the non-Markovian nature of the evolution becomes obvious. It is observed that the non-Markovian effects lead to quantum speedup in both resonance and detuning cases. As the main result of the work, it can be said that even in the  Markovian regime, increasing the detuning parameter $\delta$ leads to quantum speedup. In other words, even in the absence of memory effects, the evolution speed can be increased with the detuning parameter. It was also observed that the initial state of the system has a direct effect on the quantum speed limit time. Moreover, we have revealed that the quantum speed limit time has an inverse relation with the population of the initial excited state. Interestingly, it is also shown that the quantum speed limit time depends on the quantum coherence of the system's initial state such that the maximal coherent state can saturate the quantum speed limit bound.

\section*{ORCID iDs}
Maryam Hadipour \href{https://orcid.org/0000-0002-6573-9960}{https://orcid.org/0000-0002-6573-9960}\\
Saeed Haddadi \href{https://orcid.org/0000-0002-1596-0763}{https://orcid.org/0000-0002-1596-0763}\\
Hazhir Dolatkhah \href{https://orcid.org/0000-0002-2411-8690}{https://orcid.org/0000-0002-2411-8690}\\
Soroush Haseli \href{https://orcid.org/0000-0003-1031-4815}{https://orcid.org/0000-0003-1031-4815}

\section*{Data availability}
No datasets were generated or analyzed during the current study.


\section*{Competing interests}
The authors declare no competing interests.


\end{document}